\documentclass[fleqn,10pt]{wlscirep}
\usepackage[utf8]{inputenc}
\usepackage[T1]{fontenc}

\title{Spectral micro-CT for quantitative analysis of calcification in fibrocartilage}

\author[1,2]{Vittoria Mazzini}
\author[1,*]{Paolo Cardarelli}
\author[3]{Andrew L. Coathup}
\author[4]{Eleonora Olivotto}
\author[4]{Francesco Grassi}
\author[5]{Enrico Tassinari}
\author[1]{Simone Velardita}
\author[1,2]{Angelo Taibi}
\author[3]{Luca Brombal}
\affil[1]{INFN - Division of Ferrara, via G. Saragat 1, 44122, Ferrara, Italy}
\affil[2]{Dept. of Physics and Earth Science, University of Ferrara, via G. Saragat 1, 44122, Ferrara, Italy}
\affil[3]{Dept. of Physics, University of Trieste, and INFN - Division of Trieste, via A. Valerio 2, 34127, Trieste, Italy}
\affil[4]{RAMSES Laboratory, RIT Department, IRCCS Istituto Ortopedico Rizzoli, via di Barbiano 1/10, 40136, Bologna, Italy}
\affil[5]{2nd Orthopaedic and Traumatologic Clinic, IRCCS Istituto Ortopedico Rizzoli, via G.C.Pupilli 1, 40136, Bologna, Italy}

\affil[*]{cardarelli@fe.infn.it}


\begin{abstract}
This work introduces a quantitative method for assessing calcification in fibrocartilage using spectral micro-computed tomography (\textmu CT). Tissue samples of hip acetabular labrum from patients with osteoarthritis and femoroacetabular impingement were imaged with a laboratory-based spectral \textmu CT system equipped with a small-pixel photon-counting detector. The detector operated with two energy thresholds, allowing the simultaneous acquisition of two CT datasets at different X-ray energies. A material decomposition algorithm accounting for the system's spectral response  was applied to separate calcium- and water-like components, yielding three-dimensional visualization and quantification of calcified regions within intact paraffin-embedded samples.
Unlike the conventional method for calcification assessment based on histology, this spectral \textmu CT approach offers volumetric quantification of calcium structures without physical sectioning or staining. The method achieved a voxel size of 20 \textmu m for samples up to ~3 cm, with a calcium detection threshold of ~0.3 g/cm$^3$ for structures down to 50~\textmu m. Quantification accuracy was estimated to be 5\% by using a calibration phantom. Further comparison with histology demonstrated the correct localization of calcium spatial distributions and a match in the calcium crystal deposition score by providing non-destructive, quantitative 3D calcium maps of preserved tissue samples. This technique complements histology and could enhance the characterization of pathological fibrocartilage calcification in hip joint disorders.
\end{abstract}
\begin{document}

\flushbottom
\maketitle
\thispagestyle{empty}

\section{Introduction}

X-ray micro-computed tomography (\textmu CT) is a widely used imaging technique in biomedical research, enabling non-destructive, high-resolution, three-dimensional (3D) visualization of internal structures. Originally applied to hard tissues such as bone, \textmu CT has evolved, thanks to synchrotron radiation and advanced imaging methods, to achieve soft-tissue sensitivity and sub-cellular spatial resolution~\cite{fratini2015simultaneous,massimi2019exploring}. The latter has proven particularly valuable for detailing the 3D bone micro-architecture and characterizing mineralized tissues \cite{akhter_high_2021}. More recently, the development of laboratory-based \textmu CT systems has extended this capability beyond specialized facilities, offering a wide range of resolutions (from $\sim$100 \textmu m to $\sim$0.5 \textmu m) with scan times suitable for preclinical and ex-vivo studies~\cite{clark2021advances}. Similarly, the advent of advanced staining methods and phase-contrast imaging techniques compatible with conventional X-ray sources has extended \textmu CT of soft tissues to laboratory settings~\cite{de2019contrast,hagen2020preliminary,polikarpov2023towards,tajbakhsh2024comprehensive}. 

Photon-counting detectors (PCDs) offer a transformative alternative to EIDs. PCD can discriminate (i.e., count) individual photons by using (one or more) energy-calibrated thresholds~\cite{ballabriga2020photon}. 
This approach enhances contrast sensitivity compared with EID-based systems by suppressing electronic noise and avoiding spectral-weighting bias, without adding complexity relative to conventional scanners.
Notably, these systems enable the implementation of X-ray spectral imaging, a technique that provides material decomposition and quantitative assessment of elemental or compositional distributions~\cite{flohr2020photon,willemink2018photon}. PCDs equipped with two (or more) energy thresholds can record the incoming photons into discrete energy bins during a single exposure. By applying suitable basis material decomposition algorithms to energy-binned images~\cite{alvarez1976energy}, tissue-specific maps can be produced. These maps enable the separation and quantification of features of interest, such as a contrast agent or calcifications, from surrounding soft tissues, thereby providing richer diagnostic information. These advantages make PCDs  especially well-suited for high-precision preclinical \cite{clark2021advances} and clinical imaging~\cite{Schwartz2025}.

Recent years have seen significant advances in non-destructive imaging modalities aimed at evaluating and monitoring the health of articular cartilage and underlying subchondral bone.  High-resolution \textmu CT of ex vivo samples has emerged as a powerful technique for detailed characterization of subchondral bone morphology \cite{taheri2023changes,chappard2006subchondral}. When combined with suitable contrast agents, \textmu CT also enables volumetric assessment of cartilage structure and composition. Such contrast-enhanced studies have been carried out using conventional \textmu CT systems \cite{palmer2006analysis,stewart_synthesis_2017,bhattarai_quantitative_2018,fantoni2022cationic}, synchrotron radiation sources \cite{honkanen_triple_2020,bhattarai_dual_2020}, and more recently, compact spectral \textmu CT scanners \cite{baer_spectral_2021,fantoni2024quantitative}.
Regarding the physiopathology of osteoarthritic and pre-osteoarthritic conditions of the hip, the degree of joint tissues degeneration is conventionally assessed by histological analysis, which is a \textit{destructive approach}, as it requires processing and sectioning of the tissue samples for microscopic examination. \cite{trisolino2020labral}. 

In this work, we present a whole acquisition and processing pipeline based on spectral-\textmu CT for the quantitative characterization and analysis of calcification structures within paraffin-embedded osteoarticular samples. The method was tested on labral tissue, a fibrocartilaginous structure of the hip joint, and it is relevant for the study of pathological calcification. The samples analyzed are from patients with end-stage osteoarthritis (OA) and from patients with femoroacetabular impingement (FAI), an ex-vivo model of high-grade and low-mid grade pathological calcification of the joint, respectively. The samples, fixed in formalin and embedded in paraffin as done for conventional histology preparation \cite{trisolino2020labral}, were scanned at the PEPI Lab, using its a custom spectral \textmu CT scanner featuring a small-pixel PCD. 
The processing was performed via a two basis material decomposition, providing a quantitative estimation of the concentration of calcium and water density for each voxel. This allowed for the differentiation  and automatic segmentation of labrum components from calcified regions. The quantitative capability of the system in assessing calcium concentration derives from an accurate characterization of its spectral response, as established in previous studies~\cite{di2020characterization,brombal2022geant4}, and has been verified through experimental calibration.

Moreover, the calcium distribution maps obtained were benchmarked against histological sections of the same samples stained with Alizarin Red\texttrademark{}. This comparison indicates that the technique provides a consistent representation of the spatial distribution of calcified structures and can serve as a non-destructive approach for extracting valuable information to support cartilage grading in intact surgical samples.
\section{Methods}
\subsection{Labral samples and histological analysis\label{sec:histo}}
In this study, three labral samples were analyzed, two from patients with end-stage OA (classified with Kellgren and Lawrence - KL - grading system \cite{kellgren1957radiological}) undergoing total hip replacement and one from a FAI case undergoing arthroscopy. Samples of labral tissue from all subjects were processed for histology after fixation in 4\% formalin, dehydrated in 70\% ethanol, paraffin embedded and mounted on conventional histology cassette. As for FAI patients, surgeons were very careful in preserving tissues and avoiding labral debridement when repairing these structures is the goal. Therefore, the size of FAI sample is approximately 1 cm\textsuperscript{2} in size, while OA samples are around $\sim$4 cm\textsuperscript{2} in size. For the histological analyses, sections of 5 \textmu m were prepared and two/three sequential sections were scored. To evaluate the presence of calcification in labral samples, the sections were stained with Alizarin Red\texttrademark{} 1.4\% pH 4.2 (Sigma A5533). A \textit{calcium crystal deposition score} was assigned on a four-grade scale: grade 0 indicating no deposition, and grades 1 to 3 corresponding to increasing size and number of calcified deposits observed in the section, as previously described in Trisolino \textit{et al}\cite{trisolino2020labral}.
\subsection{\textmu CT system and samples acquisition}
The spectral \textmu CT system available at PEPI laboratory features a tungsten-anode micro-focus X-ray tube (Hamamatsu L10101) and the small-pixel PCD Pixirad-1/Pixie-III detector~\cite{bellazzini2015pixie}.  The detector has a 650~\textmu m thick CdTe sensor, thus ensuring high-efficiency up to 100~keV, and with a pixel size of 62~\textmu m that tiles a matrix of 512$\times$402~pixel. The detector features two programmable energy thresholds and an on-chip charge-sharing compensation mechanism, that mitigates cross-talk between energy bins, hence improving both spectral and spatial resolution performance~\cite{di2020characterization}.

\begin{figure}[!ht]
    \centering
    \includegraphics[width=0.95\linewidth]{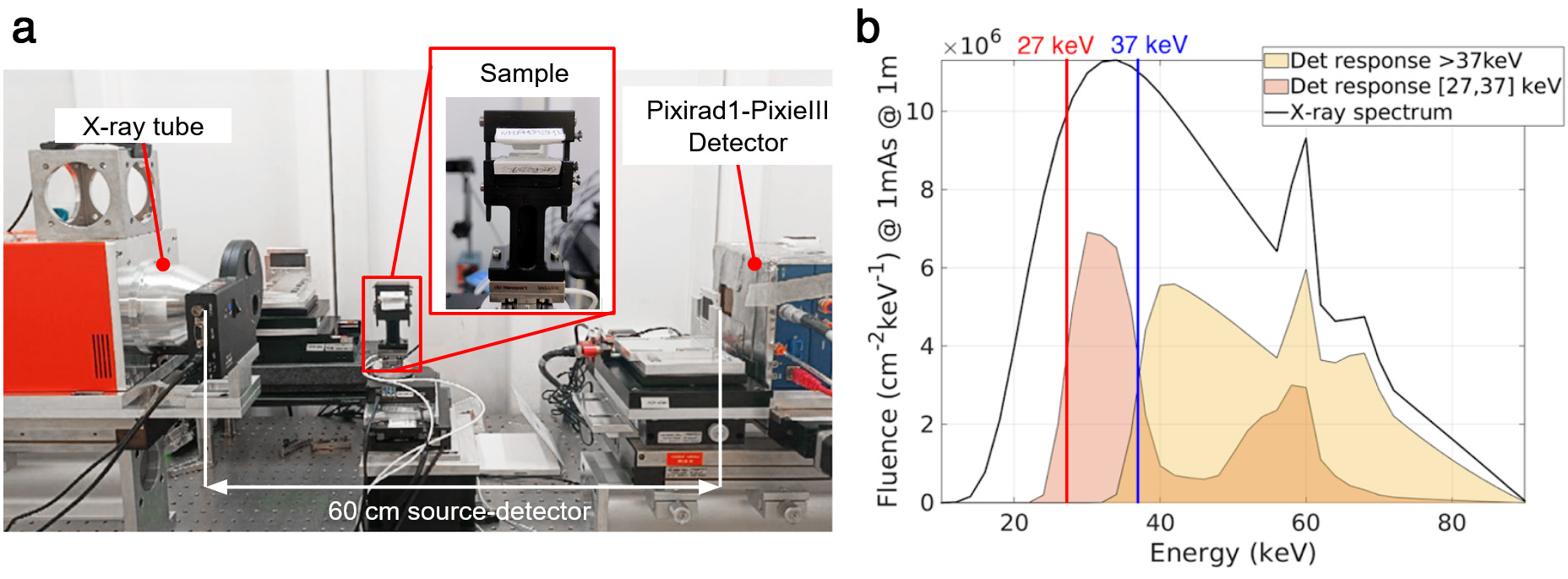}
    \caption{a) Photograph of the experimental setup. Insert shows a zoom-in of the samples mounted on the holder. b) X-ray spectrum, with detector response for the corresponding energy bins.}
    \label{fig:SETUP}
\end{figure}

The scans were performed with the X-ray tube operating at 90 kV and 150 \textmu A with a filtration of 1.75~mm of aluminum. Based on previous characterization, the used power settings correspond to a focal spot size of 19~\textmu m. The low and high energy detector thresholds were set to 27~keV and 37~keV, respectively. The low-energy threshold was selected to suppress Cd fluorescence, whereas the optimal high-energy threshold was determined using a recently developed simulation that accounts for realistic X-ray spectra, detector response, and sample interactions~\cite{Coathup2025}. The source-to-detector distance was fixed at 60~cm, while the source-to-sample distance was set to 19~cm, resulting in a geometric magnification of 3.1 and an effective voxel size of 20 \textmu m. All scans were performed in step-and-shoot mode. A schematic of the experimental setup is shown in Figure~\ref{fig:SETUP}~(a), while a plot of the used X-ray spectrum and estimated system response is reported in~Figure~\ref{fig:SETUP}~(b). A more detailed description of the system can be found in Brombal \textit{et al}\cite{brombal2023pepi}. 
As different types of samples had different lateral dimensions, acquisition parameters were chosen based on the sample at hand.
Specifically, the FAI sample (lateral dimension of $\sim$1 cm) was scanned with 1000 projections evenly distributed over 360$^\circ$. 
As the OA samples (lateral dimension of $\sim$2 cm) were larger than the detector field-of-view (FOV), they were scanned at three different detector positions. The projections corresponding to the same angle were then stitched together, allowing for the inclusion of the whole sample in the FOV~\cite{vescovi2017radiography}. Moreover, to further extend the reconstruction FOV, the axis of rotation was laterally displaced with respect to the central vertical axis of the detector~\cite{wang2002x}.
In order to ensure an adequate angular sampling, OA samples were scanned with 1500 projections over 360$^\circ$.
The projection exposure time was 10~seconds (5~repetitions of 2~seconds each) for the OA samples and 15~seconds (5~repetitions of 3~seconds each) for the FAI samples, corresponding to a total exposure time of 12.5~hours for the OA samples and of 4.2~hours for the FAI sample.
\subsection{Image processing and material decomposition}
The acquired projection images are first processed using a dedicated pre-processing algorithm~\cite{di2021pre} and subsequently reconstructed with the Feldkamp–Davis–Kress algorithm in extended FOV mode, employing the cosine reconstruction filter~\cite{wang2002x}. Each energy bin is processed independently, yielding two tomographic volumes corresponding to the low- and high-energy ranges, respectively.

A two-basis material decomposition based on the system matrix inversion is applied to the reconstructed images~\cite{alvarez1976energy}, yielding mass density maps of the selected basis materials.
Water and calcium are chosen as the basis materials, where water represents the soft tissue component and the paraffin (soft, water-like tissues), while pure calcium serves as the representative material for calcifications (hard tissues).
As detailed by Di Trapani \textit{et al}~\cite{di2022multi}, the system matrix is given by the effective mass attenuation coefficient of the basis materials evaluated over the low and high-energy bins.
The weights are given by polychromatic X-ray spectrum modulated by the detector spectral response for each energy bin, as shown in Figure \ref{fig:SETUP}~(b).
The Pixirad detector spectral response has been obtained through characterization with monochromatic radiations and dedicated Monte Carlo simulations~\cite{di2020characterization,brombal2022geant4}.
\subsection{Phantom-based calcium quantitative assessment}
In order to evaluate the accuracy of the system in the estimation of calcium density, a phantom with calibrated inserts was used. The phantom \textit{MicroCT-HA-D10} (QRM, Germany) consists of a cylindrical base made of a proprietary epoxy resin and five inserts containing different concentrations of calcium hydroxyapatite (HA), chemically represented as Ca$_5$(PO$_4$)$_3$(OH). The manufacturer provides the calibrated values of HA concentrations, as reported in Table \ref{tab:HA_phantom}, where each value is given with an uncertainty of $\pm 0.5\%$. Calcium concentration values were derived by multiplying the calibrated HA concentrations by the calcium fraction within the HA.
\begin{table}[!ht]
\centering
\begin{tabular}{lll}
Insert & HA (g/cm$^3$) & Ca (g/cm$^3$) \\
\hline
1 & 0 & 0 \\
2 & 0.049 & 0.019 \\
3 & 0.200 & 0.079 \\
4 & 0.804 & 0.321 \\
5 & 1.203 & 0.479 \\
\hline
\end{tabular}
\caption{\label{tab:HA_phantom}Certified concentrations of HA (provided by the manufacturer) and calcium.}
\end{table}
The \textmu CT acquisition of the calibration phantom was performed with the same spectrum and geometry used for the sample scans. The acquisition consisted of 1000 projections evenly distributed over 360$^\circ$, with 15 seconds exposure (5 repetitions, 3 seconds each) per projection. 
\begin{figure}[!ht]
    \centering
    \includegraphics[width=0.6\linewidth]{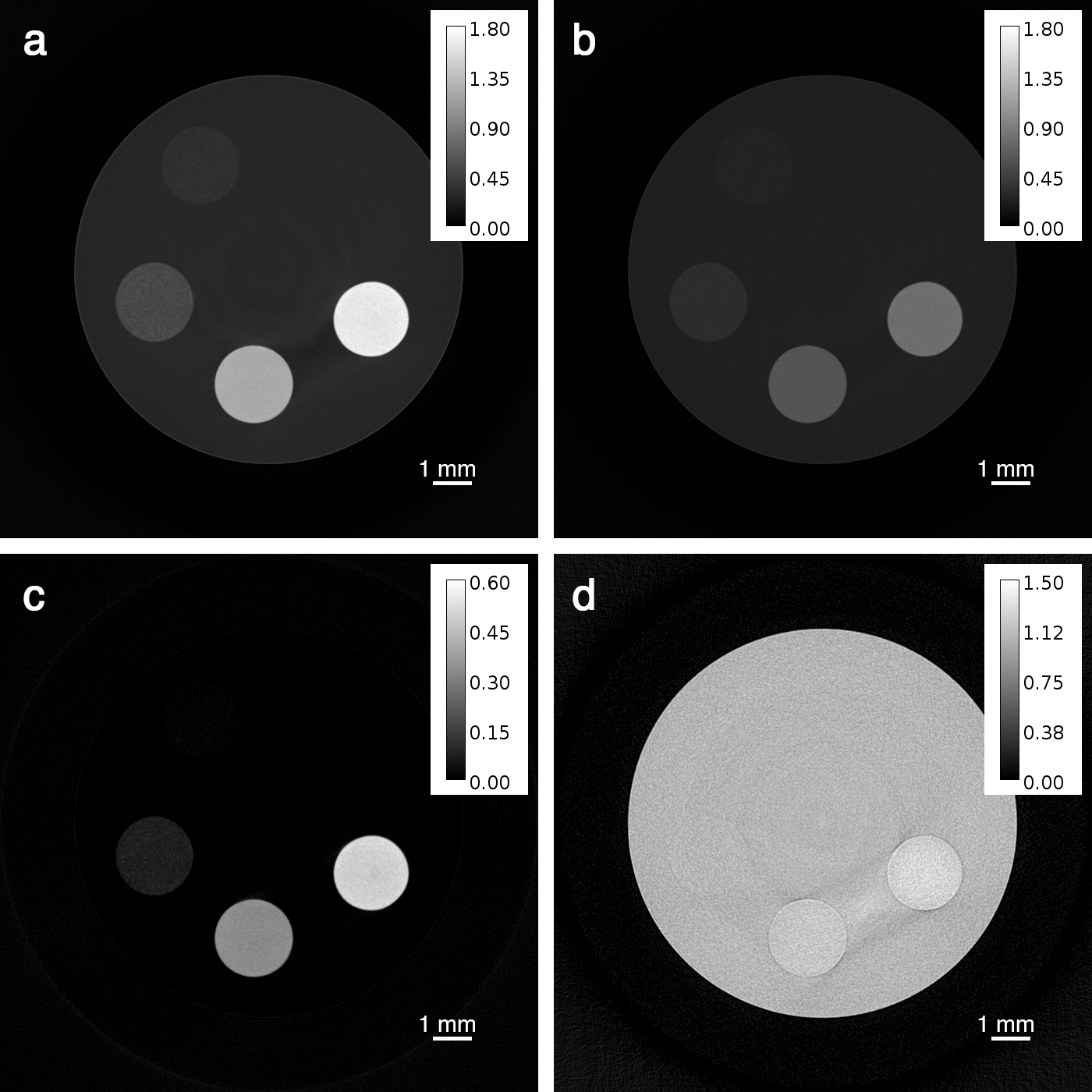}
    \caption{a) Reconstruction in the low-energy channel. Data displayed in cm\textsuperscript{-1}. b) Reconstruction in the high-energy channel. Data displayed in cm\textsuperscript{-1}. c) Decomposition in the calcium basis material. Data displayed in g/cm\textsuperscript{3}. d) Decomposition in the water basis material. Data displayed in g/cm\textsuperscript{3}.}
    \label{fig:HA_phantom}
\end{figure}
The reconstructed images in the high-energy and low-energy channels are presented in Figure~\ref{fig:HA_phantom}~(a) and~(b), while the decomposed calcium and water density maps are shown in panels~(c) and~(d), respectively. In the figure, an average of 100 consecutive slices is shown to reduce image noise. To evaluate accuracy, a region of interest (ROI) with an area of $\sim$3000 pixels was selected within each insert, and the mean and standard deviation of each ROI were measured. The measured values for the five inserts were then compared with the corresponding calibrated values listed in Table \ref{tab:HA_phantom}.
\begin{figure}[!ht]
    \centering
    \includegraphics[width=0.7\linewidth]{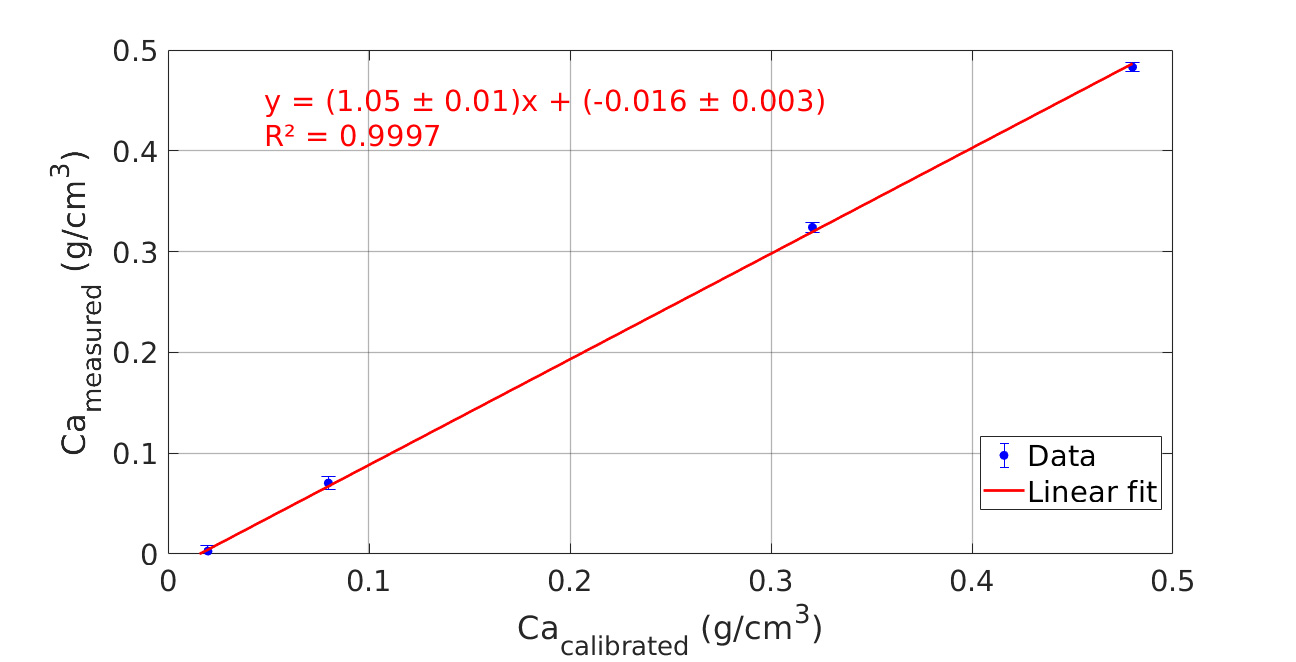}
    \caption{Measured Ca concentration versus reference values in the HA phantom.}
    \label{fig:calibration_fit}
\end{figure}
A linear fit was performed between the calibrated values presented in Table \ref{tab:HA_phantom} and the corresponding measured values obtained from the ROIs. The measured values, plotted against the reference values along with the linear fit, are shown in Figure \ref{fig:calibration_fit}. Within the typical calcium density range of the surgical samples analyzed (0.2–0.7 g/cm$^3$), measurement errors remained below 5\%, with error evaluated as the average relative discrepancy across the phantom inserts. Taking into account the linearity of the evaluation of calcium concentration (r$^2$ = 0.9997), together with the near-unity slope of the fit (1.05) and small bias (-0.016 g/cm\textsuperscript{3}), which are consistent with the mentioned 5\% uncertainty, recalibration of the absolute quantitative values was considered not necessary. The sample density in the water channel was consistently of 1.07 g/cm\textsuperscript{3} in the sample background and in the low concentrations HA inserts. Slightly higher densities (up to 1.24 g/cm\textsuperscript{3}) are associated with the highest HA concentration inserts. It should be noted that decomposition is performed in the basis materials water and calcium, whereas the phantom consists of HA and a proprietary resin with an unknown chemical composition, which accounts for the discrepancy observed in the water image.
\subsection{Calcification structures analysis}
\label{sec:analysis}
Starting from reconstructed and decomposed images, a processing pipeline aimed at the detection and characterization of calcium deposits within the labral tissue was developed. 
\begin{figure}[!ht]
    \centering
    \includegraphics[width=1.0\linewidth]{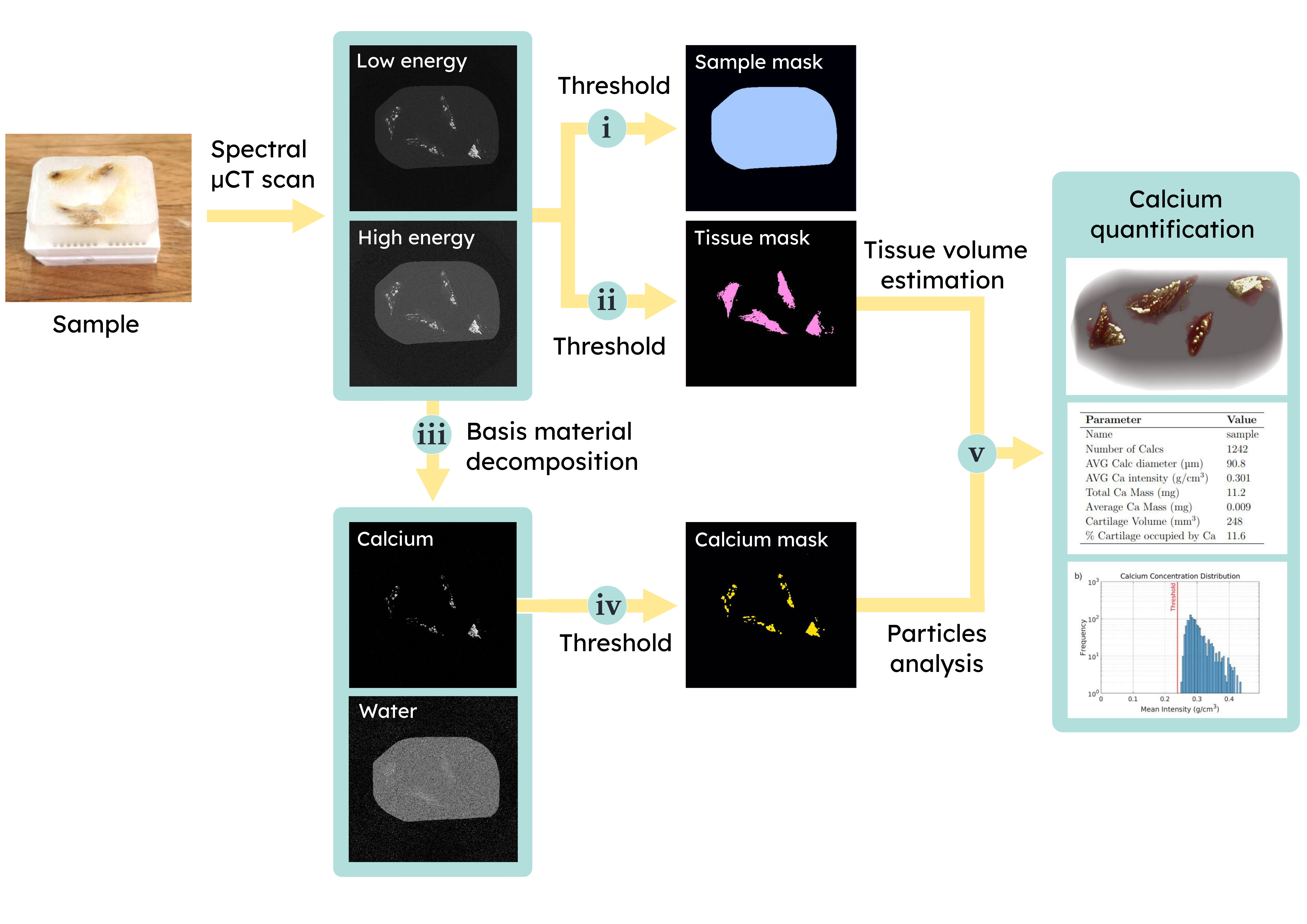}
    \caption{Schematic representation of the image processing pipeline for analyzing calcium structures in surgical samples. The numbered arrows in the diagram correspond to the steps described in Section "Calcification structures analysis".}
    \label{fig:flow}
\end{figure}
The pipeline, schematized in Figure~\ref{fig:flow}, consists of five steps hereby summarized.
\begin{enumerate}
    \item \textbf{Sample segmentation:} Sample segmentation is carried out on the low-energy reconstructions using ImageJ’s (v.~1.54g) \cite{schneider2012nih} \textit{Default} thresholding algorithm, which is a variation of the IsoData algorithm \cite{ridler1978picture}. To improve segmentation accuracy and suppress noise, a 3D median filter with a radius of 5 voxels is first applied to the low-energy images. This procedure identifies the sample volume comprising the paraffin matrix, along with the embedded cartilage and calcifications.
    
    \item \textbf{Labral tissue segmentation:} To isolate the labral tissue from surrounding paraffin, the sum of the low-energy and high-energy reconstructed images are first processed with a 3D median filter of radius 5 pixels, followed by segmentation using ImageJ’s \textit{Otsu} thresholding algorithm. The segmented labrum total volume is also quantified using \textit{MATLAB} (v.~R2023b, The MathWorks, Inc.).
   \item \textbf{Basis material decomposition:} The high and low energy reconstructed 3D images are used  as input for the two-basis material decomposition method. The results are mass density maps of the selected basis materials, water and calcium.
   
    \item \textbf{Calcium segmentation:} A thresholding method was used to isolate calcium deposits on the calcium basis image. To determine the threshold level, a ROI without calcium was selected from the calcium concentration images. This ROI was constrained by the volume of the paraffin mask, which was obtained by subtracting the labrum volume from the total sample volume. This process ensures that the selected ROI contains only paraffin, and therefore no calcifications. Within this calcium-free ROI, the mean calcium concentration ($\mu$) and standard deviation ($\sigma$) were measured to characterize the background signal distribution. A threshold value of ${TH}_{\mbox{Ca}}$ = $\mu$ + 4$\sigma$ was then applied to segment calcium from the background.
    
    \item \textbf{Calcification particles analysis:}  The sample segmentation mask is then used on the calcium map, ensuring that only calcifications within the sample are analyzed while excluding external noisy pixels that could otherwise affect the results. A custom \textit{MATLAB} script is used to process the masked decomposed images using a minimum volume threshold of 8 voxels, where this number was chosen as to avoid individual pixels behaving incorrectly. Pixels are connected if their faces, edges, or corners touch. Subsequently, the \textit{MATLAB} function \textit{regionprops3} was used to extract volumetric and intensity-based features of segmented calcium particles, including number of particles, volume, mean intensity, centroid coordinates, and maximum/minimum intensity values. Particle diameters are estimated by assuming a spherical geometry. Histograms of particle diameter and intensity distributions are generated for visualization. The labrum total volume, calculated from the segmentation in step (ii), is used for calculating the percentage of calcified volume within the tissue. Ultimately, two tables are generated: one contains detailed information for each individual particle for each sample, and another provides a summary of the entire sample.
\end{enumerate}
\section{Results}
\subsection{Calcification structure analysis}
To illustrate the methodology, results from a single OA sample (OA1) are presented, which are representative of all other samples analyzed. 

\begin{figure}[!ht]
    \centering
    \includegraphics[width=0.6\linewidth]{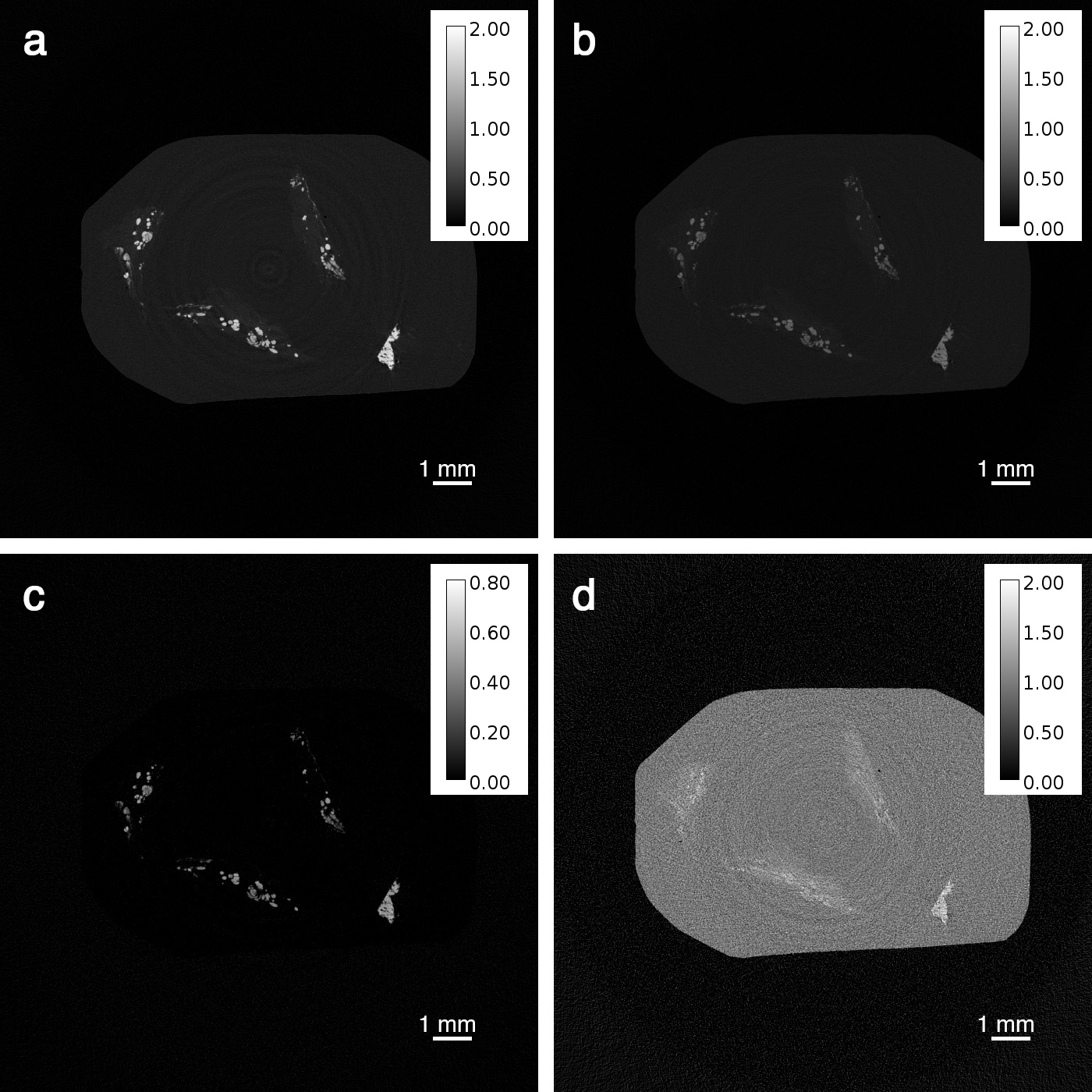}
    \caption{a) Reconstruction in the low-energy channel. Data displayed in cm\textsuperscript{-1}. b) Reconstruction in the high-energy channel. Data displayed in cm\textsuperscript{-1}. c) Decomposition in the calcium basis material. Data displayed in g/cm\textsuperscript{3}. d) Decomposition in the water basis material. Data displayed in g/cm\textsuperscript{3}.}
    \label{fig:sample}
\end{figure}

The reconstructed images of the OA1 sample in the low-energy and high-energy channels are shown in Figure~\ref{fig:sample} (a) and (b), respectively, where it is visible a marked decrease in the calcium signal going from high to low energy.
The calcium and water maps obtained via basis material decomposition are shown in Figure~\ref{fig:sample} (c) and (d), respectively. The material density maps demonstrate that spectral decomposition effectively isolates the calcified component of the sample (Figure \ref{fig:sample} (c)). Figure \ref{fig:sample} (d) shows the distribution of paraffin and labrum tissue. The signal co-localized with calcified structures is arguably due to spurious contributions in the water channel from elements present in HA other than calcium (oxygen, phosphorus).

\begin{figure}[!ht]
    \centering
    \includegraphics[width=0.7\linewidth]{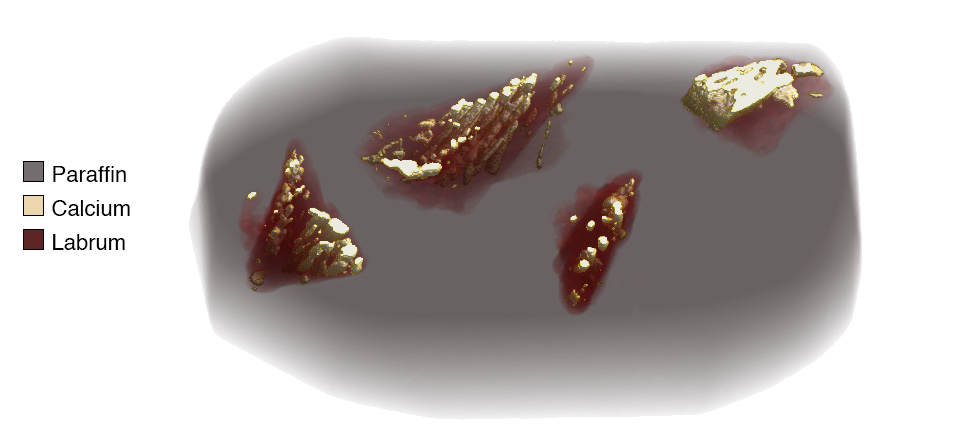}
    \caption{3D rendering of OA1 sample.}
    \label{fig:3D}
\end{figure}

\begin{figure}[!ht]
    \centering
    \includegraphics[width=0.8\linewidth]{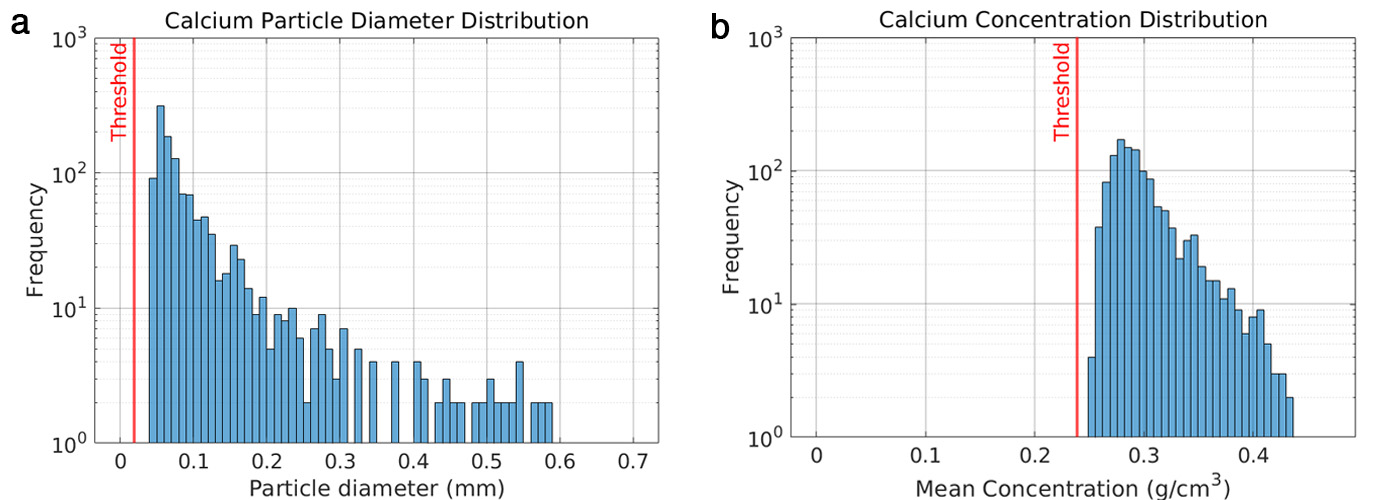}
    \caption{Results for sample OA1. a) Histogram of the Ca particles diameters distribution. b) Histogram of the mean Ca concentration for each calcification.}
    \label{fig:OA_histograms}
\end{figure}

From spectrally decomposed images, 3D renderings can be produced to improve calcification visualization, better displaying their 3D distribution. The rendering for the OA1 sample is presented in Figure~\ref{fig:3D}.
As detailed in the previous section, following particle analysis, histograms with the distributions of the particle diameter and calcium concentration can be produced, as shown for the sample OA1 in Figure~\ref{fig:OA_histograms}.
A summary of the quantitative results for all the three samples analyzed is provided in Table \ref{tab:OA_summary}. 
\begin{table}[!ht]
\centering
\begin{tabular}{llll}
\textbf{Parameter} & \textbf{OA1} & \textbf{OA2}& \textbf{FAI} \\
\hline
Number of calcifications & 1242 & 103 & 5 \\
Mean calcification diameter (\textmu m) & 123.9& 76.0& 79.4\\
Mean Ca concentration (g/cm\textsuperscript{3}) & 0.301 & 0.347 & 0.285 \\
Total Ca Mass (mg) & 11.2 & 0.013& 0.0004 \\
Average Ca Mass (\textmu g) & 8.96 & 0.13 & 0.083 \\
Cartilage Volume (mm\textsuperscript{3}) & 242& 143 & 0.418 \\
Ca/cartilage volume-fraction & 11.9\%& 0.025\% & 0.343\% \\
\hline
\end{tabular}
\caption{\label{tab:OA_summary}Summary of calcification quantitative parameters.}
\end{table}
Only calcifications with a volume greater than 8 voxel were counted, corresponding to a diameter larger than 49.6 \textmu m  in spherical approximation. The total calcium mass was calculated as the sum, over all calcifications, of the product of particle volume and mean calcium concentration. The average calcium mass was obtained as the mean of the product of particle volume and mean calcium concentration across individual calcifications. Finally, the calcium-to-cartilage volume fraction was calculated as the ratio of the total calcium volume (mean calcification volume $\times$ number of calcifications) to the total cartilage volume, expressed as a percentage. Compared to the OA samples, the calcifications in FAI ones were fewer in number, although some variability was observed across the samples. This variability likely stems from biological heterogeneity (such as patient age or disease stage).
\subsection{Comparison with histology}
\begin{figure}[!ht]
    \centering
    \includegraphics[width=0.75\linewidth]{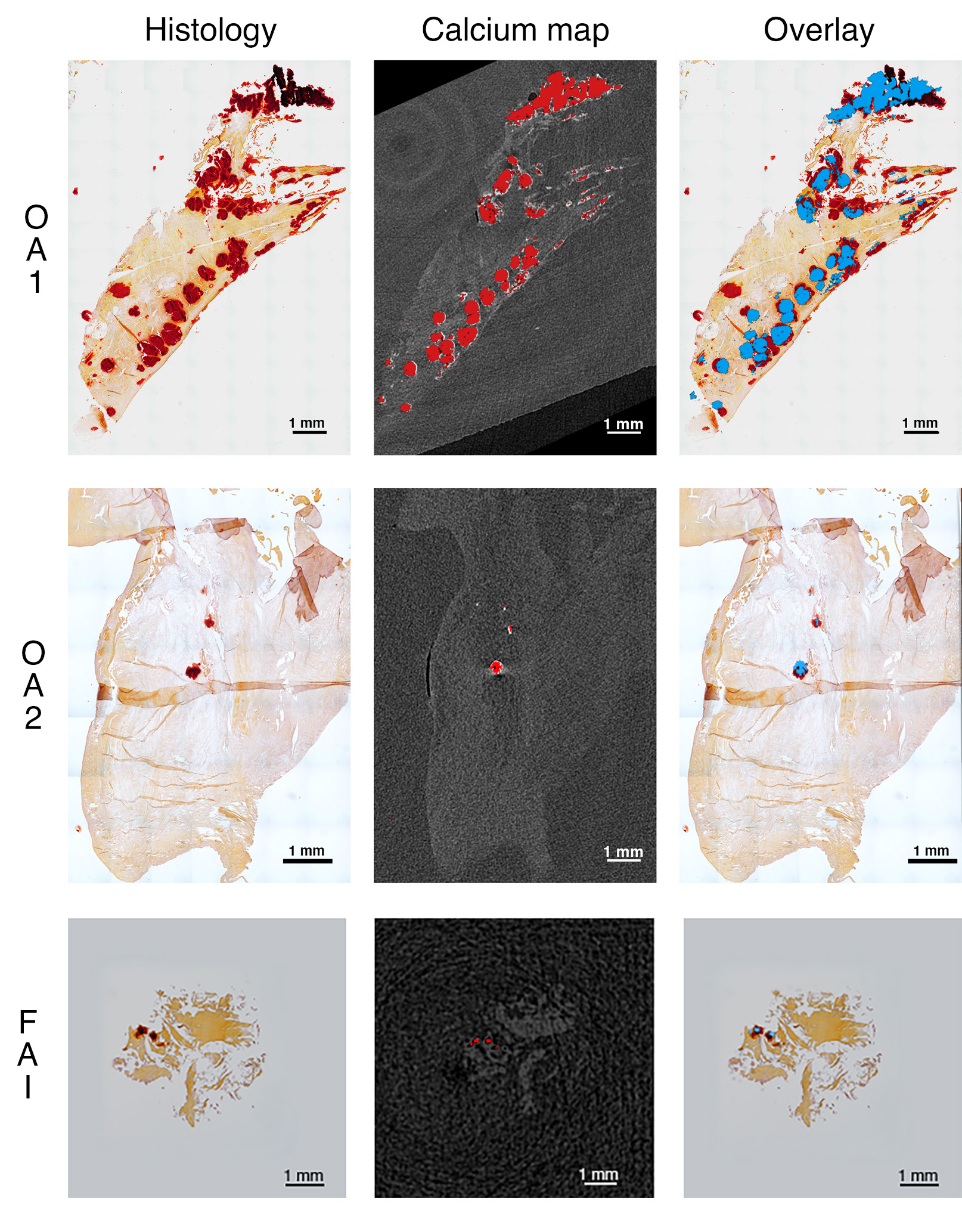}
    \caption{Comparison between histology and calcium distribution maps for both OA and FAI samples.
    \label{fig:histo_combined}}
\end{figure}

A comparison with histological analysis was performed on selected sections of the samples, to provide an independent evaluation of the spatial distribution of calcification. Sections for comparison were selected based on the presence of detectable calcium deposition in OA2 and FAI, which exhibited small calcium crystals. In contrast, OA1 showed extensive calcium deposition throughout the sample, so no specific selection was required. The results are shown in Figure \ref{fig:histo_combined} for all samples.  
The left column shows the histological section of a selected sample region, with the dark red areas corresponding to calcifications identified by Alizarin Red\texttrademark{} staining.
The center column shows the corresponding \textmu CT slice (grayscale) overlaid with the calcium distribution map (red). The right column presents the calcium distribution map as obtained by spectral \textmu CT (blue) superimposed onto the histological section. Each row corresponds to a slice of a different analyzed sample, labeled in figure. The overlay demonstrates a good spatial correspondence of calcium structures in the two modalities.

The \textit{calcium crystal deposition score}, ranging from 0 to 3, was evaluated by an expert operator based on histological analysis, also following the established conventional procedure described in Trisolino \textit{et al}\cite{trisolino2020labral}. This evaluation resulted in a score of 3 for sample OA1 and a score of 2 for samples FAI and OA2. The same grading was independently assessed for each sample using the calcium distribution images obtained with the spectral \textmu CT approach, yielding results that matched exactly those obtained from histology.
\section{Discussion}
The results demonstrate that spectral \textmu CT enables accurate non-destructive quantification of cartilage calcification in paraffin-embedded surgical samples. The setup allows reconstruction of a 3D image with a voxel size of 20 \textmu m for a field of view (FOV) up to 3.9 cm, depending on the sample size. The methodology developed allows for the detection of a minimum particle of diameter 49.6 \textmu m,  with a minimum average detectable Ca concentration of  $\sim$ 0.3 g/cm\textsuperscript{3} as determined by the threshold set at 4$\sigma$ from the Ca background signal mean value. The sample size affects the acquisition times, which ranged from approximately 4 hours to 12.5 hours, for extended FOV. This time scale is acceptable for ex-vivo studies but may limit throughput.  To address this limitation, recent advances in high-power microfocal X-ray sources, such as those based on liquid metal jet anodes, offer photon fluxes up to an order of magnitude greater than conventional X-ray tubes, enabling substantially shorter scan times~\cite{migga2022comparative}. In parallel, application-specific denoising algorithms, including neural network–based approaches, can maintain image quality while reducing acquisition time~\cite{muller2023dose,nadkarni2023deep}.

When comparing the results with the histological sections, a very good overall correspondence is observed between the images shown in Figure \ref{fig:histo_combined}, with most structures identifiable in both modalities at the same position. However, minor discrepancies are present and can be attributed to several factors. The slight spatial mismatch is due to a small difference in analysis depth. Specifically, concerning the OA1 sample, the high calcium content in the sample posed significant challenges during the slicing process, resulting in the histological cut being approximately 0.5 mm shallower than the plane analyzed via X-ray imaging. Additionally, the slices used in histology are significantly thinner (5 \textmu m)  than those in the \textmu CT scan (20 \textmu m). This leads to an uneven comparison between the two techniques. Finally, mechanical deformation of the tissue during histological preparation, particularly during cutting and mounting, may further contribute to the observed misalignment between the histological section and the X-ray data. Despite minor discrepancies in the fine spatial distribution of calcifications, the overall grading assessment remains unaffected when based on \textmu CT results. The quantification of the amount and distribution of calcified structures, showed exact agreement with the \textit{calcium crystal deposition score} based on histology of the samples.

The technique developed revealed differences between OA and FAI samples that are consistent with the distinct pathological conditions. OA specimens exhibited widespread and larger calcifications on average, in line with the chronic and degenerative nature of the disease. By contrast, FAI samples displayed only sparse and smaller calcifications, which may reflect an earlier disease stage or a primarily mechanical etiology with less extensive tissue degeneration \cite{battistelli2023hip}. Nonetheless, a systematic investigation with a larger dataset would be required to draw any conclusion regarding these observations, which is beyond the scope of the present work that is limited to the presentation and validation of the methodology.

Despite some limitations compared to conventional histology, such as a lower spatial resolution in \textmu CT and a calcium sensitivity limited by image noise (0.3 g/cm\textsuperscript{3}), there are many advantages associated with the developed approach. The technique is non-destructive, thus not interfering with any subsequent or complementary analyses of the sample. Moreover, the spectral \textmu CT approach eliminates the need for many preparatory steps, including decalcification, slicing, and staining, all of which can introduce spatial deformation and other artifacts. At the same time, unlike histology, which is limited to 2D sections, our approach provides a quantitative, fully 3D map of calcium distribution, potentially being more representative of the status of the tissue, especially for heterogeneous samples. It should be remarked that, while a simple segmentation of the calcium deposition could be obtained based solely on the absorption signal from non-spectral CT images, the proposed technique enables the quantification of the absolute calcium concentration in each voxel. This goes beyond the mere segmentation and may provide valuable insights for diagnostic purposes.
\section{Conclusions}
In this study, we presented a quantitative, automated pipeline for analyzing calcification in human acetabular labrum samples by using spectral \textmu CT imaging. By employing a small-pixel photon-counting detector and a two-material decomposition approach, we successfully distinguished and quantified calcium deposits within fibrocartilage tissues, with water and calcium as basis materials.
The method detected calcium deposits down to 49.6~\textmu m and $\sim$0.3 g/cm\textsuperscript{3} in concentration, with linearity validated against a calibration phantom and errors below 5\% in the biologically relevant range.
Furthermore, comparisons with histological analysis demonstrated a very good spatial correspondence between spectral \textmu CT and conventional methods, aside from the mismatches arising from the inherent distortions due to tissue sectioning. The calcium crystal deposition scores from histological images matched those obtained using the proposed technique for all three samples. Unlike histology, this technique is non-destructive and provides 3D maps of calcification across intact samples. This method offers a reproducible and objective alternative for assessing joint tissue calcification, with potential applications in grading osteoarticular disease, and may contribute to improving the diagnosis and investigation of joint pathologies.
\section*{Data availability statement}
The datasets generated and analyzed during the current study are available in the INFN Open Access Repository:
\\
https://doi.org/10.15161/oar.it/8ne66-f2714.
\bibliography{biblio}
\section*{Funding}
We acknowledge financial support under the National Recovery and Resilience Plan (PNRR), Mission 4, Component 2, Investment 1.1, Call for tender No. 1409 published on 14.9.2022 by the Italian Ministry of University and Research (MUR), funded by the European Union – NextGenerationEU – Project P2022X5ALY – CUP J53D23014070001- Grant Assignment Decree No. 1383 adopted on 1.9.2023 by the MUR.
\section*{Author contributions statement}
V.M. and P.C. curated the data, performed formal analysis, and prepared the original manuscript draft. P.C. additionally supervised the work, contributed to methodology, and supported funding acquisition. A. L. C. provided data curation and handled the software. E.O. and F.G. provided resources. E.T. contributed to data curation and supported manuscript review and editing. S.V. and A.T. participated in reviewing and editing the manuscript. L.B. led methodology, funding acquisition, supervision, drafting, and revision. All authors reviewed and approved the manuscript.
\section*{Ethics declarations}

\subsection*{Competing interests}
The author(s) declare no competing interests.
\subsection*{Ethical approval}
This observational study was conducted in accordance with relevant guidelines and regulations and with the approval of the Local Ethical Committee "Comitato Etico Area Vasta Emilia Centro della Regione Emilia-Romagna (CE-AVEC)" (Approval No. EM308/2024\_207/2023/Sper/IOR\_EM2). All patients were enrolled after providing informed consent.

\end{document}